\documentclass[preprint,amsmath,amssymb,aps]{revtex4-1}

\usepackage{graphicx}
\usepackage{xcolor}

\begin{document}

\title{Finsler geometric perspective on the bulk flow in the universe}

\author{Zhe Chang$^{1,2}$\footnote{E-mail: changz@ihep.ac.cn}}
\author{Ming-Hua Li$^{1}$\footnote{E-mail: limh@ihep.ac.cn}}
\author{Sai Wang$^{1}$\footnote{E-mail: wangsai@ihep.ac.cn}\footnote{Corresponding author at IHEP, CAS.}}
\affiliation{${}^1$Institute of High Energy Physics\\ Chinese Academy of Sciences, 100049 Beijing, China\\
${}^2$Theoretical Physics Center for Science Facilities\\ Chinese Academy of Sciences, 100049 Beijing, China}


\begin{abstract}
Astronomical observations showed that there may exist a bulk flow with peculiar velocities in the universe, which contradicts with the \(\Lambda\)CDM model. The bulk flow reveals that the observational universe is anisotropic at large scales. In fact, a more reliable observation on the anisotropy of spacetime comes from the CMB power spectra. The WMAP and Planck satellites both show that there is a hemispherical power asymmetry at large-angular scales. In this paper, we propose a ``wind'' scenario to the bulk flow (or the anisotropy of spacetime). Under the influence of the ``wind'', the spacetime metric could become a Finsler structure. By resolving the null geodesic equation, we obtain the modified luminosity distance, which has a dipolar form at the leading order. Thus, the ``wind'' describes well the bulk flow. In addition, we perform a least-\(\chi^2\) fit to the data of type Ia supernovae (SNe Ia) in the Union2.1 compilation. The peculiar velocity of the bulk flow has an upper limit \(v_{bulk}\lesssim 4000~\rm{km/s}\), which is compatible with all the existing observational values.
\end{abstract}
\maketitle

The cosmological principle is one of the foundations in the modern cosmology.
It assumes that the universe is statistically homogeneous and isotropic at large scales.
The principle leads to the Friedmann-Robertson-Walker (FRW) spacetime in the \(\Lambda\)CDM model \cite{Book by Dodelson}.
However, the astronomical observations recently showed that there exists a bulk flow in the universe.
The bulk flow was claimed originally by Kashlinsky {\it et al.} \cite{Kashlinsky0809,Kashlinsky0910}.
They analyzed the X-ray galaxy clusters by utilizing the kinematic Sunyaev-Zeldovich (kSZ) effect \cite{SZ1972,SZ1980}.
The bulk flow was found pointing to \((l,b)=(283^\circ\pm14^\circ,12^\circ\pm14^\circ)\)
with the peculiar velocity up to \(\sim1000~\rm{km/s}\) at the scales \(z\lesssim0.2\).
Watkins {\it et al.} \cite{Watkins0809,Feldman0911} analyzed all the peculiar velocity surveys, and found the bulk flow towards \((l,b)=(287^\circ\pm9^\circ,8^\circ\pm6^\circ)\) at the scales \(z\lesssim0.03\).
However, the peculiar velocity was found to be \(407\pm81~\rm{km/s}\), which is much smaller than Kashlinsky {\it et al.}'s results.
Furthermore, certain investigations \cite{Colin1011,Ma1010,Dai1102,Turnbull1111,Cosmological model with local symmetry of very special relativity and constraints on it from supernovae} on the type Ia supernovae (SNe Ia)
revealed the similar bulk flow as Watkins {\it et al.} claimed.
The bulk flows of such amplitudes at such large scales are not expected in the \(\Lambda\)CDM model.
The reason is that the one-dimensional r.m.s. (root mean square) velocity is expected \(\sim 110~\rm{km/s}\)
in the WMAP-normalized \(\Lambda\)CDM model \cite{Watkins0809}.

Most recently, the Planck satellite \cite{Planck2013XIII} also detected a bulk flow
extending to at least \(z\sim0.18\) by analyzing the X-ray galaxy clusters.
The amplitude of this flow is consistent with those earlier claims as mentioned above.
However, the uncertainty is much larger based on the Planck observation.
This reveals that the bulk flow is not of statistical significance.
Nevertheless, Atrio-Barandela \cite{Atrio-Barandela 1303} pointed out that
there are systematic overestimates for the uncertainty of the amplitude of bulk flow in the Planck study.
When these systematic overestimates are taken into account,
the statistical significance of the bulk flow observed by Planck satellite is consistent with those previous results.
One should note that the issue of bulk flow is indefinite until today.
The reason is that there are significant divergences on the amplitude of bulk flow between various surveys.
Though the above reported bulk flow has not been widely recognized as a fact by cosmologists and physicists,
the observations of WMAP and Planck both show that the anisotropy of spacetime cannot be neglected simply.

A more reliable observation on the anisotropy of spacetime comes from
the power spectra of CMB temperature fluctuations.
There is an anomaly in the CMB power spectra, which refers to the hemispherical power asymmetry at large-angular scales.
It was first observed at the significance \(\sim 3\sigma\)
by the WMAP observation \cite{hemispherical asymmetry01,hemispherical asymmetry02,WMAP7},
and most recently confirmed by the Planck satellite \cite{Planck 2013 resultsI,Planck2013resultsXXIII}.
The amplitude of CMB temperature fluctuations is slightly larger in one hemisphere of the sky than that in the other.
Furthermore, the hemispherical asymmetry could take the dipolar form, such as the dipolar modulation of the CMB power \cite{Spontaneous isotropy breaking}.
Recently, we have proposed that the anisotropic inflation in the Finsler spacetime
could account for the hemispherical power asymmetry \cite{Inflation and primordial power spectra
at anisotropic spacetime inspired by Planck's constraints on isotropy of CMB}.
The primordial power spectra takes the form of dipolar modulation,
which could induce the hemispherical asymmetry of CMB temperature fluctuations.
This prediction stimulates us to study the issue of bulk flow in the Finsler geometric framework.
The reason is that the bulk flow also takes the dipolar form at large scales \cite{Kashlinsky1202}.
There might be certain a relation between the hemispherical asymmetry and the large-scale bulk flow.

The existence of the bulk flow (or the hemispherical power asymmetry) implies that
the observational space is statistically anisotropic at large scales.
Thus, it contradicts with the cosmological principle.
To account for the bulk flow, we could imagine that the universe is influenced by a large-scale ``wind''.
In this way, the cosmic matter would drift with the ``wind''.
The velocity of the ``wind'' takes account for the observed peculiar velocity.
When the ``wind'' has a privileged direction, the cosmic matter should drift towards the same direction.
Thus, the large-scale bulk flow emerges in such a ``wind'' picture.
Actually, the ``wind'' picture refers to the Zermelo navigation problem \cite{Zermelo navigation problem},
which aims to find the paths of the shortest travel time in a Riemann space under the influence of a ``wind''.
It is noteworthy that the navigation problem is described by Finsler geometry \cite{Book by Bao}.
In addition, a matter dominated navigation cosmological model could account for the accelerating expansion of the universe \cite{Li1001}.

In this paper, we try to incorporate the bulk flow in the Zermelo navigation picture.
Under the influence of the ``wind'', the spatial part of the FRW metric is modified to be the Randers type \cite{Randers space}.
We study the kinematical properties of the obtained Finsler structure (line element).
The Finslerian geodesic equations are solved to obtain the relation between the cosmological redshift and the scale factor of the universe.
Further, we could obtain the modified luminosity distance.
At the first-order approximation, the luminosity distance takes a dipolar form.
This is consistent with the phenomenological estimate.
We get an upper limit on the velocity of the ``wind''
from a numerical analysis on the distance-modulus vs. redshift relation of the SNe Ia in the Union2.1 compilation \cite{Union2.1}.

First, we briefly introduce several basic issues of Finsler geometry.
Finsler geometry \cite{Book by Bao} stems from the integral of the form
\begin{eqnarray}
s=\int^b_a F\left(x, y\right)d\tau\ ,
\label{integral length}
\end{eqnarray}
where \(x\equiv x^{\mu}\) denotes a position and \(y\equiv dx /d\tau\) a so-called velocity.
The integrand $F$ is called the Finsler structure,
which is defined on the tangent bundle \(TM:=\bigcup_{x\in M}T_{x}M\) instead of the manifold \(M\).
Each element of \(TM\) is denoted by \((x,y)\) where \(x\in M\) and \(y\in T_{x}M\).
The Finsler structure is a smooth function with the positively 1-homogeneous property, i.e.,
\begin{equation}
F(x,\lambda y)=\lambda F(x,y)
\end{equation}
for all $\lambda>0$.
A manifold associated with the Finsler structure is called the Finsler manifold.
It is noteworthy that Riemann geometry belongs to Finsler geometry.

The Finsler metric tensor is defined as \cite{Book by Bao}
\begin{equation}
g_{\mu\nu}\equiv\frac{\partial}{\partial y^\mu}\frac{\partial}{\partial y^\nu}\left(\frac{1}{2}F^2\right)\ .
\end{equation}
Together with its inverse \(g^{\mu\nu}\), \(g_{\mu\nu}\) is used for lowering and raising the indices of tensors.
The parallel transport was studied via the Cartan connection \cite{Matsumoto01,Antonelli01,Szabo01}.
The length \(F\left({dx}/{d\tau}\right)\) of the vector \({dx}/{d\tau}\) is constant under the parallel transport in the Finsler spacetime.
The Finslerian geodesic equations are given by \cite{Book by Bao}
\begin{equation}
\label{geodesic}
\frac{d^2x^\mu}{d\tau^2}+G^\mu=0\ ,
\end{equation}
where
\begin{equation}
\label{geodesic spray}
G^\mu=\frac{1}{2}g^{\mu\nu}\left(\frac{\partial^2 F^2}{\partial x^\lambda \partial y^\nu}y^\lambda-\frac{\partial F^2}{\partial x^\nu}\right)
\end{equation}
denote the geodesic spray coefficients.
The equation (\ref{geodesic}) can also be deduced from the first variation of the arc integral of a regular curve \cite{Book by Bao}.

In the standard cosmological model \cite{Book by Dodelson}, the spacetime structure is described by the spatially flat FRW metric
\begin{equation}
\label{FRW metric}
ds^2=dt^2-a^2(t)\left(dx^2+dy^2+dz^2\right)\ ,
\end{equation}
where \(a(t)\) denotes the scale factor of the universe.
The cosmological redshift \(\overline{z}\) is given by
\begin{equation}
\label{redshiftinFriemann}
1+\overline{z}(t)=\frac{1}{a(t)}\ ,
\end{equation}
where we set \(a(t_{0})\equiv 1\) for today.
The evolution of the scale factor \(a(t)\) is determined by the Friedmann equation
\begin{equation}
\label{Friedmann equation}
\left(\frac{da/dt}{a}\right)^{2}=\frac{8\pi G}{3}\rho\ ,
\end{equation}
where \(\rho\) comprises the energy density of all inventory in the universe.
The luminosity distance is given by
\begin{equation}
\label{luminosity distance}
\overline{d_{L}}=\left(1+\overline{z}\right)\int_{t}^{t_{0}}\frac{dt'}{a(t')}\ ,
\end{equation}
where \(t_{0}\) denotes today.

The spatial part of the FRW metric (\ref{FRW metric}) is conformally flat.
Under the influence of a uniform ``wind'' \(W=W^{i}\frac{\partial}{\partial x^{i}}\) (\(|W|\ll 1\)),
the Zermelo navigation implies that the conformally spatial part \(\sum_{i}(dx^{i})^{2}\) could become the Randers type \cite{Li1001}
\begin{equation}
\label{Randers}
d\ell=\frac{1}{\lambda}\left(\sqrt{\lambda h_{ij}dx^{i}dx^{j}+\left(W_{i}dx^{i}\right)^{2}}-W_{i}dx^{i}\right)\ ,
\end{equation}
where \(h_{ij}=\rm{diag}(1,1,1)\) and \(\lambda=1-h_{ij}W^{i}W^{j}\).
We assume that the uniform ``wind'' is time independent.
There is a correspondence between the Riemann space influenced by the ``wind'' and the Randers space \cite{Gibbons2009}.
Thus, the Randers space can account for all the effects of the ``wind''.
It is remarkable that the above Randers space (\ref{Randers}) is flat.

By considering the temporal dimension and the Randers space (\ref{Randers}) together,
we obtain the Finslerian spacetime line element
\begin{equation}
\label{Finsler structure}
d\tau^2=dt^2-a^2(t)d\ell^2\ .
\end{equation}
Thus, the Finsler structure is
\begin{equation}
\label{F}
F=\sqrt{\left(y^{0}\right)^2-a^2(t)\left(\frac{d\ell}{d\tau}\right)^2}\ .
\end{equation}
As discussed in the following, it could account for the phenomenological estimates about the bulk flow.

In the Finsler spacetime (\ref{F}), the temporal component \(G^{0}\) of the geodesic spray coefficients is written as
\begin{equation}
\label{Temporal component}
G^{0}=\frac{1}{2}\frac{d(a^2)}{dt}\left(\frac{d\ell}{d\tau}\right)^{2}\ .
\end{equation}
For massless particles, such as the photons, the null condition \(F=0\) implies
\(a^{2}\left(\frac{d\ell}{d\tau}\right)^{2}=(y^{0})^{2}\).
The coefficient \(G^{0}\) could be rewritten as
\begin{equation}
G^{0}=\frac{\left(y^{0}\right)^{2}}{2}\frac{d\ln a^2}{dt}=\frac{y^{0}}{2}\frac{d\ln a^2}{d\tau}\ ,
\end{equation}
where \(y^{0}=dt/d\tau\).
Thus, the geodesic equation (\ref{geodesic}) could be given as
\begin{eqnarray}
\frac{d \ln\left(y^{0}\right)^{2}}{d\tau}+\frac{d\ln a^2}{d\tau}=0\ ,
\end{eqnarray}
where we have divided the whole expression by a factor \(y^{0}/2\).
The above equation has a solution
\begin{equation}
\label{solution}
ay^{0}=const.\ .
\end{equation}
The energy \(E\) of a massless particle is proportional to \(y^{0}\) (or equivalently \(a^{-1}\)).
Thus, the cosmological redshift \(z(t)\) is obtained as
\begin{equation}
\label{redshift}
1+z=\frac{1}{a(t)}\ ,
\end{equation}
which is same as the one (\ref{redshiftinFriemann}) in the FRW spacetime, i.e., \(z=\overline{z}\).

The luminosity distance could be derived from the null geodesic, i.e., \(d\tau=0\).
For simplicity, it does not lost generality to let the direction of the ``wind'' \(W\) to coincide with the third axis
in the Cartesian coordinate system.
Thus, the ``wind'' \(W\) becomes \(W=w\frac{\partial}{\partial x^{3}}\) where we have set \(W^{3}\equiv w\ll 1\).
Since the speed of ``wind'' is much smaller than the speed of light,
we could discard the effects of the ``wind'' on the azimuthal peculiar velocity.
Thus, only the radial peculiar velocity is counted in the following discussions.
This is as same as what we did in the standard cosmological model.
The null geodesic implies the relation
\begin{equation}
\label{dtdr}
\frac{dt}{a}=\frac{1}{\lambda}\left(\sqrt{\lambda+w^{2}\cos^{2}\theta}-w\cos\theta\right)dr\ ,
\end{equation}
where \(\lambda=1-w^{2}\) and \(\theta\) denotes the angle between the direction of the ``wind'' and the direction of the light.
In the observational cosmology, the luminosity distance is defined as
\begin{equation}
\label{luminosity distance in Finsler}
d_{L}\equiv\frac{r}{a(t)}=\left(1+z\right)r\ .
\end{equation}
By considering (\ref{dtdr}), we could rewrite (\ref{luminosity distance in Finsler}) as
\begin{equation}
\label{luminosity distance vs. redshift}
d_{L}=\left(1+z\right)\int^{t_{0}}_{t}\frac{dt'}{a(t')}\mathcal{A}(w,\cos\theta)\ ,
\end{equation}
where
\begin{equation}
\label{A}
\mathcal{A}\equiv\frac{\lambda}{\sqrt{\lambda+\left(w\cos\theta\right)^{2}}-w\cos\theta}\ ,
\end{equation}
and the scale factor \(a(t)\) is determined by the Friedmann equation (\ref{Friedmann equation}).
The factor \(\mathcal{A}\) accounts for all the effects of the ``wind'',
which describe the deviation of the spacetime from the FRW structure.

At the first-order approximation, we could obtain a dipolar form of \(\mathcal{A}\), i.e.,
\(\mathcal{A}\approx 1+w\cos\theta\).
Thus, the luminosity distance could be written as
\begin{equation}
\label{dipolar luminosity distance}
d_{L}=\left(1+z\right)\int^{t_{0}}_{t}\frac{dt'}{a(t')}\left(1+w\cos\theta\right)\ .
\end{equation}
By noting \(z=\overline{z}\) and substituting (\ref{luminosity distance}) into (\ref{dipolar luminosity distance}),
we find a relation between \(d_{L}\) and \(\overline{d_{L}}\) as
\begin{equation}
\label{dLdLbar}
d_{L}=\overline{d_{L}}\left(1+w\cos\theta\right)\ .
\end{equation}
This relation coincides with what used in the phenomenological analysis of the Ref.~\cite{Kashlinsky1202}.
We find that the bulk flow is indeed described by the speed of ``wind'' as we expected, i.e.,
\begin{equation}
\label{velocitywv}
v_{bulk}=w\ .
\end{equation}
This result reveals that the Finsler structure (\ref{Finsler structure}) can account for the effects of the large-scale bulk flow.
Kashlinsky {\it et al.} \cite{Kashlinsky0809,Kashlinsky0910} showed that the speed of ``wind'' is
\(\sim1000~\rm{km/s}\) towards \((l,b)=(283^\circ\pm14^\circ,12^\circ\pm14^\circ)\).
Watkins {\it et al.} \cite{Watkins0809}, however, presented a bulk flow with the velocity \(407\pm81~\rm{km/s}\)
towards \((l,b)=(287^\circ\pm9^\circ,8^\circ\pm6^\circ)\).
We note that the speed of ``wind'' differs from one analysis to the other, although the obtained directions are close to each other.

We could obtain an upper limit on the speed of ``wind'' from a simply numerical study
on the Union2.1 compilation of the SNe Ia \cite{Union2.1}.
The Union2.1 compilation contains 389 SNe Ia which have available equatorial position data
in the database of the Central Bureau for Astronomical Telegrams (CBAT) \cite{CBAT}.
We perform a least-\(\chi^2\) fit to these SNe Ia to determine the direction (\(l_0,b_0\)) and the speed \(w\) of the ``wind''.
The least-\(\chi^2\) fit is given by
\begin{equation}
\label{leastsquarefit}
\chi^2=\sum_{i=1}^{389}\frac{\left(\mu_{th}(z_i,l_0,b_0,l_i,b_i,w)-\mu_{obs}(z_i)\right)^2}{\sigma_{\mu}^{2}(z_i)}\ ,
\end{equation}
where \(\mu_{th}(z_i,l_0,b_0,l_i,b_i,w)\) is the theoretical distance modulus defined as
\begin{equation}
\label{distance modulus vs. redshift}
\mu=5\log_{10}\frac{d_{L}}{1 Mpc}+25\ .
\end{equation}
The observed distance modulus is denoted by \(\mu_{obs}(z_i)\)
while the measurement error is \(\sigma_{\mu}(z_i)\) in the Union2.1 dataset.
\(\theta_i\) denotes the angle between the unit direction of the ``wind'' and our sightline towards each SN Ia.
By performing such a least-\(\chi^2\) fit, we find no indications for the existence of such a large-scale ``wind''
at the \(95\%\) confidence level (C.L.).
Therefore, an upper limit on the speed of ``wind'' (or equivalently the bulk flow) is obtained to be \(w\lesssim 4000~\rm{km/s}\).
It is compatible with all the claims about the bulk flow.

One might wonder whether the standard general relativity could accommodate the bulk flow
by simply introducing certain corrections to the FRW metric (\ref{FRW metric}).
For instance, one may refer to the non-diagonal perturbations to the FRW metric,
such as the vector perturbation \(g_{0i}(x^{\mu})\).
As an example, we consider the simplest modification \(g_{03}=\zeta\) to the FRW metric.
Here \(|\zeta|\ll 1\) is set to be constant for simplicity.
The 0-0 component of Einstein tensor could be calculated,
which is given as
\begin{equation}
G_{00}=\frac{3\dot{a}^2}{a^2}\frac{1+5\zeta^2/3a^2}{(1+\zeta^2/a^2)^2}
-\frac{\ddot{a}}{a}\frac{2\zeta^2/a^2}{1+\zeta^2/a^2}\ .
\end{equation}
At the first order of \(\zeta/a\), it reduces back to
the conventional one \(3\dot{a}^2/a^{2}\) in the \(\Lambda\)CDM model.
However, the cosmological redshift \(z\) would acquire a first-order modification (\(\propto \zeta\)).
It could be obtained by resolving the null geodesic \(F=0\) which is calculated as
\begin{equation}
\label{null geodesic}
dt= a\left(\sqrt{1+\left(\frac{\zeta}{a}\cos\theta\right)^2}-\frac{\zeta}{a}\cos\theta\right)dr\ .
\end{equation}
Here \(\theta\) denotes the angle between our sightline and the third spatial axis.
Note that there would be significant modification to the conventional relation \(dt=adr\)
when the quantity \(\zeta/a\gg 1\).
At first order, the result for \(z\) is given as
\begin{equation}
\label{redshift for g03}
1+z=\frac{1}{a}\left(1+\zeta'\cos\theta\right)\ ,
\end{equation}
where \(\zeta'=\frac{1-a}{a}\zeta\).
We could obtain the luminosity distance (\ref{luminosity distance in Finsler}) as
\begin{equation}
d_{L}\simeq\bar{d}_{L}\left(1+v\cos\theta\right)\ ,
\end{equation}
where \(v\) denotes the mean value of  \(\frac{-{1}/{2}+\Omega_{\Lambda}/\Omega_{m}(1+z)^{3}}{1+\Omega_{\Lambda}/\Omega_{m}(1+z)^{3}}\zeta'
\approx(0.5\sim1)z\zeta\) for the SNe Ia in the Union2 data set.
In principle, there could be a bulk flow in such a modified FRW spacetime.
However, this bulk flow would contradict with the observations of the CMB temperature fluctuations today.
According to the null geodesic (\ref{null geodesic}) (or \ref{redshift for g03}),
the CMB temperature at the last scattering surface would be severely tilted with respect to the spatial directions
if there is a significant vector perturbation \(g_{03}=\zeta\) today.
In this way, the universe would be remarkably anisotropic at its primordial phase.
Nevertheless, the observations on the cosmic inhomogeneities have set severe constraints
on the level of anisotropy of the universe \cite{Dai1303}.
Thus, it is difficult for the \(g_{0i}\)-like modification to accommodate the issue of bulk flow.

Conclusions and remarks are listed as follows.
In this paper, we proposed a Finsler geometric perspective on the large-scale bulk flow in the universe.
The bulk flow was assumed to arise from a ``wind'' in the Zermelo navigation.
Under the influence of the ``wind'', the spatial part of the FRW spacetime became the Randers type.
Thus, we obtained the Finsler spacetime structure.
By studying the kinematical properties of the obtained Finsler structure,
we acquired the modified luminosity distance which is dipolar at the first-order approximation.
This prediction coincides with the phenomenological estimate.
Thus, the ``wind'' would account for the observed bulk flow as we expected.
The least-\(\chi^2\) fit was applied to the Union2.1 compilation to constrain the ``wind'' (or equivalently the bulk flow).
No indications were found for the existence of the bulk flow.
Thus, an upper limit \(v_{bulk}\lesssim 4000~\rm{km/s}\) was set on the peculiar velocity of the bulk flow.
This result is compatible with all the existing astronomical observations.

The existence of the bulk flow is contradictable with the cosmological principle.
It reveals that the universe is statistically anisotropic at large scales.
Finsler geometry is a straightforward generalization of Riemann geometry.
It is a reasonable candidate to account for the anisotropy of the spacetime.
The most significant reason is that Finsler geometry gets rid of the quadratic restriction on the spacetime structure \cite{Chern}.
In the Finsler spacetime, there are privileged axes \cite{A special-relativistic theory of the locally anisotropic space-time. I,A special-relativistic theory of the locally anisotropic space-time. II,A special-relativistic theory of the locally anisotropic space-time. Appendix,VSR in Finsler,Electromagnetic field in Finsler,Constraints on spacetime anisotropy and Lorentz violation from the GRAAL experiment,Fine structure constant variation or spacetime anisotropy,Geometrical Models of the Locally Anisotropic Space-Time}.
This could be revealed by the isometric group of the Finsler spacetime \cite{Finsler isometry by Wang,Finsler isometry by Rutz,Finsler isometry LiCM}.
For the \(d\) dimensions, there are no more than \(\frac{d(d-1)}{2}+1\) Killing vectors \cite{Finsler isometry by Wang}.
Thus, the Finsler spacetime admits less symmetries.
The Randers space is a kind of Finsler space \cite{Book by Bao}.
In general, its structure comprises the Riemann part and an extra 1-form part.
The 1-form leads the asymmetry of the Randers space under the reversal \({y^{\mu}}\longrightarrow-{y^{\mu}}\).
This property leads to a privileged axis which generates the anisotropy of the Randers space.

\vspace{0.3 cm}
\begin{acknowledgments}
We thank useful discussions with Y. Jiang, X. Li, and H.-N. Lin.
This work is supported by the National Natural Science Fund of China under Grant No. 11075166.
\end{acknowledgments}

\end{document}